\documentclass[showpacs,preprintnumbers,amsmath,amssymb,twocolumn]{revtex4}

\usepackage[normalem]{ulem}
\usepackage{graphicx}
\usepackage{dcolumn}
\usepackage{bm}
\usepackage{color}
\newcommand{\n}{\nonumber}
\newcommand{\p}{\partial}
\newcommand{\la}{\langle}
\newcommand{\ra}{\rangle}
\newcommand{\rad}{\ra_{\cal D}}
\newcommand{\dv}{\Delta\varphi}

\begin{document}

\title{Relative information entropy and Weyl curvature of the inhomogeneous
Universe}

\author{Nan Li~\footnote{Electronic address: \texttt{linan@mail.neu.edu.cn}}}
\affiliation{Department of Physics, College of Sciences, Northeastern University, Shenyang 110819, China}
\author{Thomas Buchert~\footnote{Electronic address:
\texttt{buchert@obs.univ-lyon1.fr}}}
\affiliation{Universit\'e de Lyon, Observatoire de Lyon, Centre de Recherche Astrophysique de Lyon, CNRS UMR 5574: Universit\'e Lyon~1
and \'Ecole Normale Sup\'erieure de Lyon, 9 avenue Charles Andr\'e, F-69230 Saint-Genis-Laval, France}
\author{Akio Hosoya~\footnote{Electronic address:
\texttt{ahosoya@th.phys.titech.ac.jp}}} \affiliation{Department of Physics, Tokyo Institute of Technology, Oh-Okayama, Meguro, Tokyo 152-8551, Japan}
\author{Masaaki Morita~\footnote{Electronic address:
\texttt{morita@okinawa-ct.ac.jp}}} \affiliation{Department of Integrated Arts and Science, Okinawa National College of Technology,
905 Henoko, Nago, Okinawa 905-2192, Japan}
\author{Dominik J. Schwarz~\footnote{Electronic address:
\texttt{dschwarz@physik.uni-bielefeld.de}}}
\affiliation{Fakult\"at f\"ur Physik, Universit\"at Bielefeld, Universit\"atsstra\ss e 25, Bielefeld D-33615, Germany}

\begin{abstract}
Penrose conjectured a connection between entropy and Weyl curvature
of the Universe. This is plausible, as the almost homogeneous and
isotropic Universe at the onset of structure formation has
negligible Weyl curvature, which then grows (relative to the Ricci
curvature) due to the formation of large-scale structure, and thus
reminds us of the second law of thermodynamics. We study two scalar
measures to quantify the deviations from a homogeneous and isotropic
space-time: the relative information entropy and a Weyl tensor
invariant, and show their relation to the averaging problem. We
calculate these two quantities up to second order in standard
cosmological perturbation theory and find that they are correlated
and can be linked via the kinematical backreaction of a spatially
averaged universe model.
\end{abstract}


\maketitle

\section{Introduction} \label{sec:intro}

The starting point of cosmology is commonly based on the simple
Friedmann-Lema\^{\i}tre-Robertson-Walker (FLRW) model, which
describes a homogeneous and isotropic space-time, as a mathematical
realization of the strong cosmological principle. However, due to
gravitational instability, our Universe obviously deviates from this
simple model locally, and hosts various large-scale structures,
e.g., the anisotropies in the cosmic microwave background. Spatial
homogeneity and isotropy may be valid only on scales larger than
hundreds of Mpc, and the effects due to inhomogeneities within these
scales are in the focus of many investigations. While the
conventional approach assumes that the FLRW model also provides a
valid description {\it on average} of the inhomogeneous Universe, in
general we expect structure formation to affect the average
evolution: the FLRW model is in general not only locally, but also
globally gravitationally unstable~\cite{phasespace}. As a first
step, we shall here confront average properties of inhomogeneous
models with a perturbative modeling of inhomogeneities using
standard perturbation theory.

Penrose conjectured that some scalar invariant of the Weyl curvature
tensor is a monotonically growing function of time and could be
identified with the gravitational entropy of the
Universe~\cite{Penrose1,Penrose2}. This conjecture, also known as
the ``Weyl curvature hypothesis'', to our knowledge was neither
formulated in a rigorous way, nor has the precise notion of
``entropy'' been specified. However, it is very plausible that some
entropy measure exists, as a FLRW space-time has vanishing Weyl
curvature and non-vanishing Ricci curvature, whereas Black Hole
solutions of space-time have nonzero Weyl and (for the simplest
models) zero Ricci curvatures. As it is assumed that the Universe
evolves from an almost homogeneous and isotropic space-time towards
an ensemble of randomly distributed Black Holes in the far future,
the Weyl curvature seems to grow relative to the Ricci curvature of
the Universe. A monotonically increasing function reminds us of the
second law of thermodynamics, so Penrose wondered whether the Weyl
tensor somehow describes or is related to some notion of
``gravitational entropy'' of the Universe. An attempt to define the
gravitational entropy has been made for a linearly perturbed FLRW
solution, taking tensor perturbations as purely gravitational
ones~\cite{Mena2007}.

We propose in this paper to identify the notion of ``entropy'' with
an information theoretical measure. On a finite patch of the
Universe such a measure can be defined, once one has specified an
averaging proceedure. In the context of cosmology, such an averaging
procedure was proposed by one of the present
authors~\cite{buchert:onaverage1,buchert:onaverage2}, and it has
been applied to the Kullback-Leibler information entropy by some of
the present authors in Ref.~\cite{entropy}. Here we combine those
findings with cosmological perturbation theory, motivated by the
Weyl curvature hypothesis.

In Sect.~\ref{sec:measure}, we first introduce two scalar measures
to quantify inhomogeneities in the Universe: the relative
information entropy and a Weyl tensor invariant, and next show that
they both can be related to the averaging problem in a perturbed
universe model in Sect.~\ref{sec:ACPT}. In
Sect.~\ref{sec:calculation}, we calculate these two quantities by
means of cosmological perturbation theory up to second order and
prove that they can be correlated via the kinematical backreaction
of a spatially averaged universe model in Sect.~\ref{sec:relation}.
We conclude in Sect.~\ref{sec:conclusion}.

\section{Measures of inhomogeneity in the perturbed Universe} \label{sec:measure}

Since our Universe is regionally inhomogeneous and anisotropic, the
value of a physical observable $O(t,{\bf x})$ at a specific time and
place is of little significance to us, and only its average $\la
O(t,{\bf x})\rad$ over a domain $\cal D$ is meaningful and practical
in cosmology. The importance of the averaging problem in the
perturbed Universe lies in the fact that many cosmological
observables are averaged quantities. The most obvious example is the
Hubble constant $H_0$. We pick $N$ galaxies in a local volume $V$
and measure their luminosity distances $d_i$ and recession
velocities $v_i$, and $H_0=\frac{1}{N}\sum_{i=1}^N\frac{v_i}{d_i}$.
Therefore, in the limit of a big sample, it naturally turns into a
volume average $H_0=\frac{1}{V}\int\frac{v}{d}\,d^3x$. (We should
state that cosmological information is generally encoded on the past
light-cone, so the averaging problem should be done over a
light-cone volume~\cite{Marozzi1,Marozzi2}. However, for objects
with redshifts $\ll 1$, spatial averaging on a
hypersurface of constant time is already a good approximation, because
the Hubble rate does not change significantly on the temporal scale
involved.)

There are many ways to quantify the deviation of $O(t,{\bf x})$ from
$\la O(t,{\bf x})\rad$. For instance, the variance ${\rm
var}(O):=\la(\delta O)^2\rad=\la(O-\la O\rad)^2\rad=\la O^2\rad-\la
O\rad^2$ is a good choice. In this section, we are going to present
two different methods to describe this deviation: one is the
Kullback-Leibler relative entropy in information theory, and the
other is the Weyl tensor in differential geometry. They both can
characterize the degree of inhomogeneity of the perturbed Universe.

\subsection{Kullback-Leibler relative information entropy} \label{sec:KL}

The relative information entropy was introduced by Kullback and
Leibler~\cite{KL} first in information theory as
\begin{eqnarray}
S\{p||q\}:=\sum_{i}p_i\ln\frac{p_i}{q_i}, \n 
\end{eqnarray}
where $\{p_i\}$ and $\{q_i\}$ denote the actual and presumed
probability distributions, respectively. It is easy to see that this
entropy vanishes if the two distributions agree, and will be
positive definite if the actual distribution departs from the
presumed one. Therefore, the relative information entropy may
quantify the divergence from the actual distribution to the
theoretical one, which is usually used in an idealized model such as
an assumed background.

This idea was generalized to the study of the evolution of an
inhomogeneous Universe by Hosoya, Buchert, and Morita. In
Ref.~\cite{entropy}, a straightforward and reasonable extension of
the Kullback-Leibler relative information entropy from a discrete
system to a continuum in cosmology was derived from the basic
principle of non-commutativity of the two operations ``spatial
averaging'' and ``temporal evolution'':
\begin{eqnarray}
&&\frac{{\cal S}_{\cal D}\{\rho||\la\rho\rad\}}{V_{\cal D}}:=\left\la\rho\ln\frac{\rho}{\la\rho\rad}\right\rad, \quad {\rm with} \n\\
&&\la\rho\dot{\rad}-\la{\dot\rho}\rad=-\frac{\dot{\cal S}_{\cal D}}{V_{\cal D}}\;. \label{entropy}
\end{eqnarray}
Here, ${\cal S}_{\cal D} / V_{\cal D}$ is the relative information
entropy density in a domain $\cal D$ with the volume $V_{\cal D}$,
and $\rho$ and $\la\rho\rad$ are the actual and averaged energy
densities in the Universe, as the analogues of $\{p_i\}$ and
$\{q_i\}$. (For a general discussion of the Kullback-Leibler
relative information entropy and its applications in different
models in cosmology, we refer to Refs.~\cite{M,Akerblom:2010mx}.)

\subsection{Weyl tensor} \label{sec:Weyl}

The Weyl tensor is a measure of the curvature of a pseudo-Riemannian
manifold. In four-dimensional space-time, it is defined as
\begin{eqnarray}
C_{\mu\nu\lambda\rho}&:=&R_{\mu\nu\lambda\rho}+\frac12(g_{\mu\rho}R_{\nu\lambda}+g_{\nu\lambda}R_{\mu\rho}-g_{\mu\lambda}R_{\nu\rho} \n\\
&&-g_{\nu\rho}R_{\mu\lambda})+\frac{1}{6}(g_{\mu\lambda}g_{\nu\rho}-g_{\mu\rho}g_{\nu\lambda})R,\label{Weyl}
\end{eqnarray}
where $R_{\mu\nu\lambda\rho}$ and $R_{\mu\nu}$ are the coefficient
functions of the Riemann and Ricci tensors, and
$R_{\mu\nu}:={R^\lambda}_{\mu\lambda\nu}$. Our sign convention is based on
the metric signature ($-,+,+,+$). The Weyl tensor has
the same symmetries as the Riemann tensor:
$C_{\mu\nu\lambda\rho}=-C_{\nu\mu\lambda\rho}=-C_{\mu\nu\rho\lambda}$,
$C_{\mu\nu\lambda\rho}=C_{\lambda\rho\mu\nu}$, and
$C_{\mu\nu\lambda\rho}+C_{\mu\lambda\rho\nu}+C_{\mu\rho\nu\lambda}=0$.
But the Weyl tensor is traceless, i.e., the contraction on any pair
of indices yields zero, e.g., ${C^\lambda}_{\mu\lambda\nu}=0$. So
the contraction of the Weyl tensor
$C_{\mu\nu\lambda\rho}C^{\mu\nu\lambda\rho}$ is the principal scalar
invariant we can construct. (Strictly speaking, there is another
principal scalar invariant
$^{\star}C_{\mu\nu\lambda\rho}C^{\mu\nu\lambda\rho}$, with
$^{\star}C_{\mu\nu\lambda\rho}$ being the left dual of the Weyl
tensor, but in this paper we restrict our exploration to
$C_{\mu\nu\lambda\rho}C^{\mu\nu\lambda\rho}$ for simplicity.)

The Weyl tensor possesses an important property: if the Weyl tensor
vanishes, the metric of space-time is locally conformally flat. This
means that we may transform the metric tensor to a constant tensor
in a local coordinate system by a conformal transformation. Two
metric tensors are said to be conformally equivalent, if
$\tilde{g}_{\mu\nu}(\tilde{x})=\Omega^2(x)g_{\mu\nu}(x)$. In this
circumstance, they have the same Weyl tensor, and their null
geodesics coincide. (For an introduction to conformal transformation
and conformal equivalence, see Ref.~\cite{Dabrowski:2008kx}.)

We assume that, at early times, when the Universe is still almost
homogeneous and perturbations have not grown significantly, the
metric of space-time can be approximately represented by the
zero-curvature FLRW one, $ds^2=-dt^2+a^2(t)\delta_{ij}dx^idx^j$. By
a conformal transformation, $dt=a(t)d\eta$, we obtain
$ds^2=a^2(\eta)(-d\eta^2+\delta_{ij}dx^idx^j)$. Therefore, the
metric is conformally equivalent to a flat Minkowski one, and its
Weyl tensor vanishes automatically. Whereas, in the late Universe,
space-time becomes highly inhomogeneous, and its metric departs from
the FLRW one and is thus not conformally flat any more.
Hence, the Weyl tensor appears in the perturbed Universe.

Penrose conjectured an analogue~\cite{Penrose1,Penrose2} of the
emergence of the Weyl tensor to the entropy increasing in
thermodynamics of Black Holes. For a Black Hole with mass $M$, the
Schwarzschild metric is
$ds^2=-(1-\frac{2GM}{r})dt^2+(1-\frac{2GM}{r})^{-1}dr^2+r^2d\Omega^2$,
and $C_{\mu\nu\lambda\rho}C^{\mu\nu\lambda\rho}=48(GM)^2/r^6$.
Meanwhile, the entropy $S$ of the Schwarzschild Black Hole is
$S=\frac{k_{\rm B}}{4G\hbar}A=\frac{k_{\rm B}}{4G\hbar}\times
4\pi(2GM)^2$, with $A=4\pi(2GM)^2$ being its area. We observe
clearly that $C_{\mu\nu\lambda\rho}C^{\mu\nu\lambda\rho}$ is
proportional to $S$, so there may be some latent relation of the
Weyl tensor to the entropy of Black Holes and gravitational fields.

In cosmology, during the process of structure formation, the cosmic
structures decouple from the global expansion of the Universe little by
little and become gravitationally bound systems. Generally speaking,
these systems will eventually end their evolution as Black Holes.
Consequently, Black Holes and their corresponding Weyl tensors will
together arise here and there in the Universe, and the average of
$C_{\mu\nu\lambda\rho}C^{\mu\nu\lambda\rho}$ can thus be considered
as another measure of structure formation, or another kind of
entropy in some sense. But we should point out that Penrose's
conjecture is not a well formulated mathematical statement, and the
aim of this paper is to explore its relation to the relative
information entropy and other relevant physical quantities.

\vskip .3cm

In this section, we have discussed two different methods to measure
the degree of inhomogeneity of the Universe from the unperturbed
background: one is the relative entropy in information theory, and
the other is the Weyl tensor in differential geometry. They root
from different branches of science and are seemingly uncorrelated at
first look. However, besides these apparent distinctions, they have
essential points in common. First, they are both related to the
entropy increasing in the evolution of the Universe; second, and
more importantly, they are both related on the level of spatially
averaged quantities in the inhomogeneous Universe, as we shall
demonstrate. Therefore, in the next sections, we shall explain the
averaging problem in the perturbed Universe, calculate
$\mathcal{S}_{\cal D}$ and $\la C_{\mu\nu\lambda\rho}C^{\mu\nu\lambda\rho}\rad$ in cosmological
perturbation theory, and finally show their relation in Sect.~\ref{sec:relation}.

\section{Averaging problem and cosmological perturbation theory} \label{sec:ACPT}

In this section, we first recall the averaging procedure in an
inhomogeneous Universe together with the effective Friedmann
equations for an irrotational dust universe model. The matter model
``irrotational dust'' is assumed throughout the paper and implied
when we talk about ``cosmic continuum''. Next, we turn to
cosmological perturbation theory and provide all the necessary
mathematical preparations for Sects.~\ref{sec:calculation} and
\ref{sec:relation}, the main part of this paper.

\subsection{Averaging problem in cosmology} \label{averaging}

The essence of the averaging problem in cosmology lies in the
non-commutation of spatial averaging and temporal evolution. As a
consequence, inhomogeneities and anisotropies influence the
evolution of the background (averaged) universe model. Let us start
from a compact domain $\cal D$ with metric perturbations at initial
time $t_{\rm i}$. If we first smooth out the fluctuations of the
metric at $t_{\rm i}$, what remains is merely a simple FLRW model.
Afterwards, the evolution of this averaged model to time $t$ is
nothing but a pure expansion. In contrast, if we exchange the order
of these two operations: first follow the evolution of the perturbed
space-time from $t_{\rm i}$ to $t$, and then take the average of
physical observables in the resulting domain at $t$, we clearly
arrive at a different result. This mechanism is usually named as
cosmological backreaction~\cite{Ellis,ellisbuchert}. (See
Ref.~\cite{Buchert} for a comprehensive review of the backreaction
formalism, and Ref.~\cite{BR} for a recent overview.)

To understand the averaging problem mathematically, we start from
the kinematics of the perturbed Universe. The covariant derivative
of the four-velocity of the cosmic continuum can be decomposed as
$u_{\mu;\nu}=\frac13h_{\mu\nu}\theta+\sigma_{\mu\nu}+\omega_{\mu\nu}$,
where $h_{\mu\nu}$ is the projection operator,
$\sigma^{\mu}_{~\nu}:={h^{\mu}}_{\alpha}{h^{\beta}}_{\nu}\left[\frac{1}{2}({u^{\alpha}}_{;\beta}+{u_{\beta}}^{;\alpha})
-\frac{1}{3}{h^{\alpha}}_{\beta}{u^{\lambda}}_{;\lambda}\right]$ is
the shear tensor, $\sigma^2:=\frac12{\sigma^{\mu}}_{\nu}{\sigma^{\nu}}_{\mu}$ is the
shear scalar, and $\theta:={u^{\lambda}}_{;\lambda}$ is the
expansion scalar. In the following, the vorticity tensor
$\omega_{\mu\nu}$ has to vanish by construction in a flow-orthogonal
foliation of space-time; we assume that this is a good approximation
on large scales, since vorticity decays in perturbation theory due
to the expansion of the Universe. Moreover, we neglect radiation at
the late times, and use the dust approximation. Last, in the
following, we perform the concrete calculation for the simplest
Einstein-de Sitter dust model, in which the cosmological constant is
supposed to vanish. (It is reasonable because the Einstein-de Sitter
model with small perturbations is believed to be an excellent
approximation in most of the matter-dominated epoch, during which
our calculation applies. A generalization with the cosmological
constant is straightforward, but that will not change our following
statements in a qualitative way.)

Below, we follow the averaging formalism proposed in
Refs.~\cite{buchert:onaverage1,buchert:onaverage2}. The metric of
the perturbed Universe can be written in the synchronous gauge as
$ds^2=-dt^2+g_{ij}(t, {\bf x})\,dx^idx^j$, and the spatial
Riemannian volume average of an observable $O(t,\bf x)$ in a
comoving domain $\cal D$ at time $t$ is defined as
\begin{eqnarray}
\langle O \rad:=\frac{1}{V_{\cal D}(t)}\int_{\cal D} O(t,{\bf x})\sqrt{\mbox{det}g_{ij}}\,d^3x, \n
\end{eqnarray}
with $V_{\cal D}(t):=\int_{\cal D}\sqrt{{\rm det}g_{ij}}\,d^3x$
being the Riemannian volume of $\cal D$. An effective volume scale
factor, $a_{\cal D}(t)/a_{\cal D}(t_0):=(V_{\cal D}(t)/V_{\cal
D}(t_0))^{1/3}$, can thus be introduced. For later evaluation within
perturbation theory, we define the Euclidian average $\la O\ra:=\int
O(t,{\bf x})\,d^3x/\int\,d^3x$, i.e., the integral without
$\sqrt{\mbox{det}g_{ij}}$ on the background cosmology.

By averaging the energy constraint
and the Raychaudhuri equation, we arrive at the effective Friedmann
equations for the irrotational dust universe model (the Buchert
equations)~\cite{buchert:onaverage1,buchert:onaverage2},
\begin{eqnarray}
\left(\frac{\dot{a}_{\cal D}}{a_{\cal D}}\right)^2+\frac{k_{\cal D}}{a_{\cal D}^2}&=&\frac{8\pi G}{3}\la\rho\rad-\frac{{\cal Q}_{\cal D}+{\cal W}_{\cal D}}{6},\n\\
\frac{\ddot{a}_{\cal D}}{a_{\cal D}}&=&-\frac{4\pi G}{3}\la\rho\rad+ \frac{{\cal Q}_{\cal D}}{3}, \n\\
\la\rho\dot{\rad} + 3 \frac{\dot{a}_{\cal D}}{a_{\cal D}}\la\rho\rad&=& 0 .\n
\end{eqnarray}
In these equations, we observe that besides the ordinary entries in
the Friedmann equations for the FLRW model, there appear two
additional terms: the deviation of the averaged spatial curvature
from a constant-curvature model, ${\cal W}_{\cal D}: = \langle {\cal
R}\rangle_{\cal D} - 6 k_{\cal D} / a_{\cal D}^2$, and the
kinematical backreaction ${\cal Q}_{\cal D}$,
\begin{eqnarray}
{\cal Q}_{\cal D}:=\frac{2}{3}(\la\theta^2\rad-\langle\theta\rad^2)-2\langle\sigma^2\rad.\label{Q}
\end{eqnarray}
These two terms are not independent and can be linked via an
integrability condition $(a_{\cal D}^6{\cal Q}_{\cal
D}\dot{)}+a_{\cal D}^4(a_{\cal D}^2 {\cal W}_{\cal D}\dot{)}=0$. In
the following, we shall be using a background model with $k_{\cal D}= 0$
so that the variable ${\cal W}_{\cal D}$ is equal to $\la {\cal R}\rad$.

\subsection{Cosmological perturbation theory} \label{CPT}

For the perturbative calculations of the relative information
entropy and the Weyl tensor invariant in
Sect.~\ref{sec:calculation}, we briefly list the basics of
cosmological perturbation theory. Here we concentrate on the scalar
modes at the first order for a flat irrotational dust model.

In the comoving synchronous gauge, the perturbed metric of
space-time reads
\begin{eqnarray}
ds^2=-dt^2+a^2(t)\left[(1-2\Psi)\delta_{ij}+D_{ij}\chi\right]dx^idx^j.\label{metric}
\end{eqnarray}
Here the scale factor $a(t)=(t/t_0)^{2/3}$ ($a(t_0):=1$) is slightly
different from the effective scale factor $a_{\cal D}(t)$, and their
relation was provided in Ref.~\cite{Li1}. $\Psi(t,{\bf x})$ and
$\chi(t,{\bf x})$ are the scalar metric perturbations,
$D_{ij}:=\p_i\p_j-\frac13\delta_{ij}\Delta$, and $\Delta$ is the
three-dimensional Laplace operator. Substituting this perturbed
metric into the Einstein equations, we obtain the solutions for
$\Psi$ and $\chi$ (only the growing modes are taken into
account)~\cite{Li1,Li2},
\begin{eqnarray}
\Psi(t,{\bf x})&=&\frac{1}{2}\Delta\varphi ({\bf x})t_0^{4/3}t^{2/3}+\frac{5}{3}\varphi({\bf x}), \n\\
\chi(t,{\bf x})&=&-3\varphi({\bf x})t_0^{4/3}t^{2/3}, \label{solution}
\end{eqnarray}
where $\varphi({\bf x})$ is the time-independent
peculiar-gravitational potential, defined from the Poisson equation,
$\Delta\varphi({\bf x}):=4\pi Ga^2(t)\rho^{(1)}(t,{\bf x})$.
Similarly, we may directly find the energy density of the background
universe model $\rho^{(0)}(t)$ and the perturbation
$\rho^{(1)}(t,{\bf x})$ over it,
\begin{eqnarray}
\rho^{(0)}(t)=\frac{1}{6\pi Gt^2}, \quad \rho^{(1)}(t,{\bf x})=\frac{1}{4\pi
G}\left(\frac{t_0}{t}\right)^{4/3}\Delta\varphi({\bf x}).\label{rho}
\end{eqnarray}

We should emphasize that in the following sections we shall perform
all the perturbative calculations up to second order. However, here
we only mention the first order results, but this is already enough,
as we shall see immediately that both the relative information
entropy and contraction of the Weyl tensor are second order
quantities and do not involve zeroth order quantities, and can thus
be constructed by using only the first order perturbations. All this
will be carefully explained in Sects.~\ref{pSS} and \ref{pW}~\cite{Chris}.

\vskip .3cm

The above discussion completes all the mathematical preparations for
the forthcoming perturbative calculations.

\section{Perturbative calculations of the relative information entropy and the Weyl tensor invariant} \label{sec:calculation}

We now move on to the detailed perturbative calculations of the
behavior of the relative information entropy ${\cal S}_{\cal D}$ and
the spatial average of the contraction of the Weyl tensor $\la
C_{\mu\nu\lambda\rho}C^{\mu\nu\lambda\rho}\rad$ in the evolution of
the perturbed Universe.

\subsection{Perturbative calculation of the relative information entropy} \label{pSS}

At the onset of structure formation, when energy density
perturbations are small, we expand $\rho$ to second order, and the
relative information entropy density is
\begin{eqnarray}
\frac{{\cal S}_{\cal D}}{V_{\cal D}}&=&\left\la{\rho\ln\frac{\rho}{\la\rho\rad}}\right\rad \n\\
&=&\left\la(\rho^{(0)}+\rho^{(1)}+\rho^{(2)})\times\right. \n\\
&&\left.\ln\left(\frac{\rho^{(0)}+\rho^{(1)}+\rho^{(2)}}{\rho^{(0)}+\la\rho^{(1)}\rad+\la\rho^{(2)}\rad}\right)\right\rad \n\\
&=&\frac{1}{2}\frac{\la(\rho^{(1)})^2\ra-\la\rho^{(1)}\ra^2}{\rho^{(0)}}+{\rm higher~order~terms}.\label{pS}
\end{eqnarray}
We see from Eq.~(\ref{pS}) that, although the second order
perturbation $\rho^{(2)}$ enters the expression of ${\cal S}_{\cal
D} / V_{\cal D}$, it does not explicitly enter into the final result
at second order. This is understandable, as the leading term of the
entropy density is the variance of the energy density, which is
already a second order term. Moreover, at second order, we are
entitled to use the Euclidian average $\la\cdots\ra$ to approximate
$\la\cdots\rad$, since their differences are terms of even higher
order~\cite{Chris}.

Substituting the solution of $\rho^{(1)}$ from Eq.~(\ref{rho})
into Eq.~(\ref{pS}), we obtain
\begin{eqnarray}
\frac{{\cal S}_{\cal D}}{V_{\cal D}}&=&\frac{3}{16\pi G}\frac{t_0^{8/3}}{t^{2/3}}[\la(\Delta\varphi)^2\ra-\la\Delta\varphi\ra^2] \n\\
&=&\frac{3}{16\pi G}\frac{t_0^{8/3}}{t^{2/3}}{\rm var}(\dv)\propto\frac{1}{t^{2/3}}.\label{Sdensity}
\end{eqnarray}
We find clearly from Eq.~(\ref{Sdensity}) that ${\cal S}_{\cal D}$
is positive, which is consistent with its definition. Since $V_{\cal
D}\propto a_{\cal D}^3\propto a^3\propto t^2$, we observe from
Eq.~(\ref{Sdensity}) that
\begin{eqnarray}
{\cal S}_{\cal D}\propto V_{\cal D}^{2/3}\propto a_{\cal D}^2\propto a^2\propto t^{4/3}. \label{S}
\end{eqnarray}
Equation~(\ref{S}) shows that ${\cal S}_{\cal D}$ increases in the
evolution of the perturbed Universe, and thus it indeed is a measure
that characterizes the degree of structure formation, as we expect.

Furthermore, the results for ${\dot{\cal S}}_{\cal D} / V_{\cal D}$
and ${\ddot{\cal S}}_{\cal D} / V_{\cal D}$ can directly be
obtained,
\begin{eqnarray}
\frac{{\dot{{\cal S}}}_{\cal D}}{V_{\cal D}}&=&\frac{4}{3t}\frac{{\cal S}_{\cal D}}{V_{\cal
D}}=\frac{1}{4\pi G}\frac{t_0^{8/3}}{t^{5/3}}{\rm var}(\dv), \label{SS1}\\
\frac{{\ddot{{\cal S}}}_{\cal D}}{V_{\cal D}}&=&\frac{4}{9t^2}\frac{{\cal S}_{\cal D}}{V_{\cal D}}
=\frac{1}{12\pi G}\frac{t_0^{8/3}}{t^{8/3}}{\rm var}(\dv).\label{SS2}
\end{eqnarray}
We find that both of them are positive, meaning that ${\cal S}_{\cal
D}$ not only increases monotonically, but in an accelerated manner.

The general result for the evolution of the relative information
entropy has a profound relation to the non-commutation of spatial
averaging and temporal evolution in the averaging problem. It was
proved in Ref.~\cite{entropy} that
\begin{eqnarray}
\frac{{\dot{{\cal S}}}_{\cal D}}{V_{\cal D}}
=\la\dot{\rho}\rad-\la\rho\rad^{^{\textbf{.}}}=\la\rho\rad\la\theta\rad-\la\rho\theta\rad=-\la\delta\rho\delta\theta\rad \; .\n \\
\end{eqnarray}
Therefore, in the process of structure formation, whether for an
overdense or underdense region, we both have
\begin{enumerate}
\item Overdense region: $\delta\rho>0$ and $\delta\theta<0$,
$-\la\delta\rho\delta\theta\rad>0$, so ${\dot{{\cal S}}}_{\cal D}>0$.
\item Underdense region: $\delta\rho<0$ and $\delta\theta>0$,
$-\la\delta\rho\delta\theta\rad>0$, so ${\dot{{\cal S}}}_{\cal D}>0$.
\end{enumerate}
Thus, generally speaking, ${\cal S}_{\cal D}$ increases
monotonically, and this is in agreement with our linear perturbative
results.

Last, we can prove that the time convexity of relative information
entropy $\frac{{\ddot{{\cal S}}}_{\cal D}}{V_{\cal D}}$ is
\begin{eqnarray}
\frac{{\ddot{{\cal S}}}_{\cal D}}{V_{\cal D}}&=&4\pi G{\rm var}(\rho)+\frac{1}{3}\la\rho(\delta\theta)^2\rad+2\la\rho\sigma^2\rad \n\\
&&+\la\rho\rad{\cal Q}_{\cal D}-\frac{2}{3}\frac{{\dot{{\cal S}}}_{\cal D}}{V_{\cal D}}\la\theta\rad. \n
\end{eqnarray}
Unfortunately, it is a highly nontrivial task to figure out the sign
of this exact result. But we see from Eq.~(\ref{SS2}) that in a
perturbative approach, up to second order, ${\ddot{{\cal S}}}_{\cal
D} / V_{\cal D}$ is still positive.

At this level of calculation, we can safely assume that the relative
information entropy increases monotonically during the evolution of
the Universe, and it thus serves as a measure of structure
formation.

\subsection{Perturbative calculation of the Weyl tensor invariant} \label{pW}

Starting from the perturbed metric in Eq.~(\ref{metric}),
\begin{eqnarray}
ds^2=-dt^2+a^2(t)[(1-2\Psi)\delta_{ij}+D_{ij}\chi]\,dx^idx^j,\n
\end{eqnarray}
the calculation for the contraction of the Weyl tensor
$C_{\mu\nu\lambda\rho}C^{\mu\nu\lambda\rho}$ is straightforward,
although a little bit tedious. Here we show some intermediate steps
before giving the final result. Due to the symmetry of the Weyl
tensor, 112 of its 256 components vanish automatically, and all the
remaining ones are the derivatives of a function $C$, a combination
of the metric perturbations $\Psi$ and $\chi$,
\begin{eqnarray}
C:=\Psi+\frac{1}{6}\Delta\chi-\frac{1}{2}\left(a\dot{a}\dot{\chi}+a^2\ddot{\chi}\right).\n
\end{eqnarray}
For example, some typical components are
\begin{eqnarray}
&&C_{0101}=\frac12D_{11}C, \quad C_{0102}=\frac12D_{12}C, \n\\
&&C_{1212}=\frac{a^2}{2}(D_{11}+D_{22})C, \quad C_{1213}=\frac{a^2}{2}D_{23}C. \n
\end{eqnarray}
Therefore, direct calculation shows
\begin{eqnarray}
&&C_{\mu\nu\lambda\rho}C^{\mu\nu\lambda\rho} \n\\
&=&\frac{1}{a^4}\{2[D_{11}CD_{22}C+D_{22}CD_{33}C+D_{33}CD_{11}C] \n\\
&&~~~~+3[(D_{11}C)^2+(D_{22}C)^2+(D_{33}C)^2] \n\\
&&~~~~+4[(D_{12}C)^2+(D_{23}C)^2+(D_{31}C)^2]\}.\n
\end{eqnarray}

In a dust universe model, from the solutions in Eq.~(\ref{solution}), we
have $C=2\varphi$, so
\begin{eqnarray}
C_{\mu\nu\lambda\rho}C^{\mu\nu\lambda\rho}&=&8\left(\frac{t_0}{t}\right)^{8/3}\left[\partial^i(\partial_j\varphi\partial^j\partial_i\varphi)\right. \n\\
&&\left. -\partial^i(\partial_i\varphi\Delta\varphi)+\frac{2}{3}(\Delta\varphi)^2\right],\n
\end{eqnarray}
and
\begin{eqnarray}
\la C_{\mu\nu\lambda\rho}C^{\mu\nu\lambda\rho}\rad&=&8\left(\frac{t_0}{t}\right)^{8/3}\left[\la\p^i(\p_j\varphi\partial^j\partial_i\varphi)\ra\right. \n\\
&&\left. -\la\partial^i(\partial_i\varphi\Delta\varphi)\ra+\frac{2}{3}\la(\Delta\varphi)^2\ra\right].\label{CC}\n \\
\end{eqnarray}
Equation~(\ref{CC}) implies, similarly to ${\cal S}_{\cal D}/V_{\cal
D}$, that $\la C_{\mu\nu\lambda\rho}C^{\mu\nu\lambda\rho}\rad$ is
also a second order quantity. Since all the components of
$C_{\mu\nu\lambda\rho}$ are zero at the FLRW background, i.e. there
are no product terms of second order perturbations with (here
vanishing) zeroth order terms, we only need the linear
perturbation theory to calculate $\la
C_{\mu\nu\lambda\rho}C^{\mu\nu\lambda\rho}\rad$ up to second order.

\vskip .3cm

Equations~(\ref{Sdensity}), (\ref{SS1}), (\ref{SS2}), and (\ref{CC})
display all the perturbative results for the relative information
entropy and the Weyl tensor invariant. However, their relation seems
still unclear, but we shall prove immediately that they actually can
be related via the kinematical backreaction term ${\cal Q}_{\cal D}$.

\section{Relation of the kinematical backreaction} \label{sec:relation}

In this section, we shall finally show the relation of the relative
information entropy and the Weyl tensor invariant with the
kinematical backreaction term ${\cal Q}_{\cal D}$,
\begin{eqnarray}
{\cal Q}_{\cal D}:=\frac{2}{3}(\la\theta^2\rad-\langle\theta\rad^2)-2\langle\sigma^2\rad.\n
\end{eqnarray}
${\cal Q}_{\cal D}$ consists of two parts: the variance of the
expansion scalar $\theta$ and the variance of the shear scalar
$\sigma$ (where the average of $\sigma$ itself could be added to the
backreaction term and accordingly added to the kinematical part of
the equations, which then would feature Bianchi-type kinematics).
Since both of them are second order terms, we can again calculate
them with only the first order perturbed metric in
Eq.~(\ref{metric}). Detailed perturbative calculations can be found
in Refs.~\cite{Li1,Li2}, and here we only list the necessary
results,
\begin{eqnarray}
\theta&=&\frac{\dot{a}}{a}-3\dot{\Psi}, \quad {\sigma^i}_j={\theta^i}_j-\frac{1}{3}\theta{\delta^i}_j=\frac{1}{2}{D^i}_j\dot{\chi},\n\\
\sigma^2&=&\frac{1}{2}{\sigma^i}_{j}{\sigma^{j}}_i=\frac{1}{8}{D^i}_j\dot{\chi}{D^j}_i\dot{\chi} \n\\
&=&\frac{t_0^{8/3}}{2t^{2/3}}\left[\partial^i\partial_j\varphi\partial^j\partial_i\varphi-\frac{1}{3}(\Delta\varphi)^2\right]. \n
\end{eqnarray}
(These results may be understood along the following lines: the Weyl
curvature can be irreducibly split into the electric part
$E_{\mu\nu}$ and the magnetic part $H_{\mu\nu}$. If only linear
scalar perturbations are treated, the magnetic part vanishes, so the
Weyl curvature squared $C_{\mu\nu\lambda\rho}C^{\mu\nu\lambda\rho}$
should be proportional to the electric part squared $E_{\mu\nu}
E^{\mu\nu}$. Also in the linear approximation, the shear tensor is
proportional to the electric part, and consequently
$C_{\mu\nu\lambda\rho}C^{\mu\nu\lambda\rho}$ is proportional to the
shear squared $\sigma^2$. This fact is easily seen by comparing
$C_{\mu\nu\lambda\rho}C^{\mu\nu\lambda\rho}$ in Eq.~(\ref{CC}) and
$\sigma^2$ above.) Therefore, up to second order, we have
\begin{equation}
{\cal Q}_{\cal D}=\frac{t_0^{8/3}}{t^{2/3}}\left[\langle\partial^i(\partial_i\varphi\Delta\varphi)\ra
-\la\partial^i(\partial_j\varphi\partial^j\partial_i\varphi)\rangle-\frac{2}{3}\langle\Delta\varphi\rangle^2\right]. \label{pQ}
\end{equation}
We also remark that, due to the vanishing of the magnetic part, and
due to the fact that in this perturbative approach the perturbations
are propagating on the background, this result is formally
equivalent to the corresponding Newtonian
result~\cite{ehlersbuchert:weyl}.

Taking Eq.~(\ref{pQ}), we eventually find that the difference
between the relative information entropy ${\cal S}_{\cal D} /V_{\cal
D}$ in Eq.~(\ref{Sdensity}) and the contraction of the Weyl tensor
$\la C_{\mu\nu\lambda\rho}C^{\mu\nu\lambda\rho}\rad$ in
Eq.~(\ref{CC}) is nothing but the kinematical backreaction term
${\cal Q}_{\cal D}$. These three quantities are correlated in the
following relation (up to second order),
\begin{eqnarray}
\frac{{\cal S}_{\cal D}}{V_{\cal D}}=\frac{9}{32\pi G}\left(\frac{t^2}{8}\la
C_{\mu\nu\lambda\rho}C^{\mu\nu\lambda\rho}\rad+{\cal Q}_{\cal D}\right).\label{result}
\end{eqnarray}
We may get further insight from the result in Eq.~(\ref{result}),
especially for the limit in which the volume of the averaging domain
$V_{\cal D}$ goes to zero. From Eqs.~(\ref{CC}) and (\ref{pQ}), we
observe that $\la C_{\mu\nu\lambda\rho}C^{\mu\nu\lambda\rho}\rad$
and ${\cal Q}_{\cal D}$ are almost the same, except for the
difference, $\la(\Delta\varphi)^2\ra-\langle\Delta\varphi\rangle^2$,
i.e., the variance of the energy density perturbation $\rho^{(1)}$
(see Eq.~(\ref{rho})). So in the limit $V_{\cal D}\rightarrow 0$,
the averaging effects from inhomogeneities turn out to be
negligible, and we should recover in a nice way the point-wise
defined parameters for the unperturbed Universe. Therefore,
$\langle(\Delta\varphi)^2\rangle$ and $\la\Delta\varphi\rangle^2$
cancel each other, and the relative information entropy ${\cal
S}_{\cal D}$ thus reduces to zero, as it should be for the
coincidence of the actual distribution with the assumed one. The
above argument provides a self-consistency check of the averaging
framework~\cite{WS}.

The content of our relation, Eq.~(\ref{result}), may also be
restated by using the relative entropy production density
${\dot{\cal S}}_{\cal D} / V_{\cal D}$ of Eq.~(\ref{SS1}),
\begin{eqnarray}
\frac{{\dot{{\cal S}}}_{\cal D}}{V_{\cal D}}=\frac{3}{8\pi G}\left(\frac{t}{8}
\la C_{\mu\nu\lambda\rho}C^{\mu\nu\lambda\rho}\rad+\frac{{\cal Q}_{\cal D}}{t}\right),\label{resultd}
\end{eqnarray}
or by using ${\ddot{{\cal S}}}_{\cal D} /V_{\cal D}$ of
Eq.~(\ref{SS2}),
\begin{eqnarray}
\frac{{\ddot{{\cal S}}}_{\cal D}}{V_{\cal D}}=\frac{1}{8\pi G}\left(\frac{1}{8}\la
C_{\mu\nu\lambda\rho}C^{\mu\nu\lambda\rho}\rad+\frac{{\cal Q}_{\cal D}}{t^2}\right).\label{resultdd}
\end{eqnarray}
We should point out that the sign of ${\cal Q}_{\cal D}$ is not
explicit in perturbation theories, and is even more difficult to
determine in exact solutions. At last we should state that, even
though the results in Eqs.~(\ref{result}) - (\ref{resultdd}) are
obtained from the perturbative calculations in the synchronous
gauge, they are gauge invariant, because all the three quantities
${\cal S}_{\cal D}/V_{\cal D}$, $\la
C_{\mu\nu\lambda\rho}C^{\mu\nu\lambda\rho}\rad$, and ${\cal Q}_{\cal
D}$ are second order ones and vanish at both the zeroth (background)
and first orders~\cite{Stewart}.

\section{Conclusions} \label{sec:conclusion}

The study of inhomogeneous models in relativistic cosmology and the
relevant averaging problems have attracted more and more attention
in recent
years~\cite{Buchert,ellis:focus,clarkson:review,kolb:focus,rasanen:focus,wiltshire:focus,buchert:focus},
among which one of the central issues is to seek a simple and
reasonable measure of the inhomogeneous distribution of the cosmic
continuum. In this paper, we first discuss two such measures as an
assessment of the degree of structure formation in the perturbed
Universe: the relative information entropy ${\cal S}_{\cal D}$ and
the contraction of the Weyl tensor $\la
C_{\mu\nu\lambda\rho}C^{\mu\nu\lambda\rho}\rad$; we calculate these
two seemingly uncorrelated quantities in the standard perturbative
approach up to second order. We find that both of them are related
to the general averaging problem in the inhomogeneous Universe and
show their relation via the kinematical backreaction term ${\cal Q}_{\cal D}$.

Eqs.~(\ref{result}) - (\ref{resultdd}) are the main results of our
paper. From these relations, we prove that ${\cal S}_{\cal D}/V_{\cal D}$ is proportional to $\la
C_{\mu\nu\lambda\rho}C^{\mu\nu\lambda\rho}\rad$, iff ${\cal Q}_{\cal
D}$ is negligible, i.e., iff the averaged Universe is described by an
idealized FLRW cosmology. However, since kinematical backreaction
measures the deviation from a FLRW cosmology, an improved background
including backreaction could eventually result in a direct
proportionality, if perturbations on the general average are
considered, rather than on a FLRW background. The answer to this
question is beyond the scope of this paper, but the formalism to
address this more general issue has been outlined in a recent
work~\cite{roy:perturbations}.

Regarding the Weyl curvature hypothesis we can state that it is not
the Weyl curvature alone that is monotonically growing, but a
specific combination with the kinematical backreaction, a result that
has been proved here to second order in perturbation theory. In
other words, the information entropy ${\cal S}_{\cal D}$ consists of
the gravitational entropy $\la C_{\mu\nu\lambda\rho}C^{\mu\nu\lambda\rho}\rad$, proposed by
Penrose, and the kinematical backreaction ${\cal Q}_{\cal D}$, which
contributes to the observed cosmic acceleration. It will be
interesting to find out if this could be extended to the fully
non-linear theory.

\vskip .3cm

We are very grateful to Yun-He Li, Bo-Qiang Ma, and Xin Zhang for
fruitful discussions, to Xavier Roy and Alexander Wiegand for
valuable comments on the manuscript. The work of NL is supported by
the National Natural Science Foundation of China (No. 11105026). NL
and DJS thank the Deutsche Forschungsgemeinschaft (DFG) for
financial support via the grant IRTG 881. TB's work was conducted
within the ``Lyon Institute of Origins'' under grant ANR-10-LABX-66.


\begin{thebibliography}{99}

\bibitem{phasespace}
X. Roy, T. Buchert, S. Carloni, and N. Obadia: Global gravitational
instability of FLRW backgrounds -- Interpreting  the dark sectors.
\emph{Class. Quant. Grav.} \textbf{28}, 165004 (2011) [arXiv:1103.1146 [gr-qc]].

\bibitem{Penrose1}
R. Penrose: Singularities and time asymmetry. In {\it General
relativity, an Einstein centenary survey}, edited by S. W. Hawking
and W. Israel (Cambridge, Cambridge University Press, 1979) p. 581.

\bibitem{Penrose2}
R. Penrose: Before the big bang: an outrageous new perspective and
its implications for particle physics. In \emph{Proceedings of EPAC 2006}, (2006) p. 2759.
http://accelconf.web.cern.ch/accelconf/e06/PAPERS/
THESPA01.PDF.

\bibitem{Mena2007}
F. C. Mena and P. Tod: Lanczos potentials and a definition of
gravitational entropy for perturbed FLRW space-times. \emph{Class.
Quant. Grav.} \textbf{24}, 1733 (2007) [arXiv:gr-qc/0702057].

\bibitem{buchert:onaverage1}
T. Buchert: On average properties of inhomogeneous fluids in general
relativity: dust cosmologies. \emph{Gen. Rel. Grav.} \textbf{32},
105 (2000) [arXiv:gr-qc/9906015].

\bibitem{buchert:onaverage2}
T. Buchert: On average properties of inhomogeneous fluids in general
relativity: perfect fluid cosmologies. \emph{Gen. Rel. Grav.}
\textbf{33}, 1381 (2001) [arXiv:gr-qc/0102049].

\bibitem{entropy}
A. Hosoya, T. Buchert, and M. Morita: Information entropy in
cosmology. \emph{Phys. Rev. Lett.} \textbf{92}, 141302 (2004)
[arXiv:gr-qc/0402076].

\bibitem{Marozzi1}
M. Gasperini, G. Marozzi, F. Nugier, and G. Veneziano: Light-cone
averaging in cosmology: formalism and applications. \emph{JCAP} {\bf
1107} 008 (2011) [arXiv:1104.1167 [astro-ph.CO]].

\bibitem{Marozzi2}
I. Ben-Dayan, M. Gasperini, G. Marozzi, F. Nugier, and G. Veneziano:
Backreaction on the luminosity-redshift relation from gauge
invariant light-cone averaging. \emph{JCAP} {\bf 1204}, 036 (2012)
[arXiv:1202.1247 [astro-ph.CO]].

\bibitem{KL}
S. Kullback and R. A. Leibler: On information and sufficiency.
\emph{Ann. Math. Stat.} {\bf 22}, 79 (1951).

\bibitem{M}
M. Morita, T. Buchert, A. Hosoya, and N. Li: Relative information
entropy of an inhomogeneous Universe. \emph{AIP Conf. Proc.} {\bf
1241}, 1074 (2010) [arXiv:1011.5604 [gr-qc]].

\bibitem{Akerblom:2010mx}
N. Akerblom and G. Cornelissen: Relative entropy as a measure of
inhomogeneity in general relativity. \emph{J. Math. Phys.} {\bf 53},
012502 (2012) [arXiv:1008.5288 [gr-qc]].

\bibitem{Dabrowski:2008kx}
M. P. D\c{a}browski, J. Garecki, and D. B. Blaschke: Conformal
transformations and conformal invariance in gravitation.
\emph{Annalen Phys.} {\bf 18}, 13 (2009) [arXiv:0806.2683 [gr-qc]].

\bibitem{Ellis}
G. F. R. Ellis: Relativistic cosmology: its nature, aims and
problems. In \textit{General relativity and gravitation}, edited by
B. Bertotti, F. de Felice, and A. Pascolini (D. Reidel Publishing
Company, Dordrecht, 1984), p. 215.

\bibitem{ellisbuchert}
G. F. R. Ellis and T. Buchert: The Universe seen at different
scales. \emph{Phys. Lett. A.} (Einstein Special Issue) \textbf{347},
38 (2005) [arXiv:gr-qc/0506106].

\bibitem{Buchert}
T. Buchert: Dark energy from structure: a status report. \emph{Gen.
Rel. Grav.} \textbf{40}, 467 (2008) [arXiv:0707.2153 [gr-qc]].

\bibitem{BR}
T. Buchert and S. R\"as\"anen: Backreaction in late-time cosmology.
\emph{Ann. Rev. Nucl. Part. Sci.} \textbf{62}, 57 (2012)
[arXiv:1112.5335 [astro-ph.CO]].

\bibitem{Li1}
N. Li and D. J. Schwarz: On the onset of cosmological backreaction.
\emph{Phys. Rev. D} {\bf 76}, 083011 (2007) [arXiv:gr-qc/0702043].

\bibitem{Li2}
N. Li and D. J. Schwarz: Scale dependence of cosmological
backreaction. \emph{Phys. Rev. D} {\bf 78}, 083531 (2008)
[arXiv:0710.5073 [astro-ph]].

\bibitem{Chris}
C. Clarkson, K. Ananda, and J. Larena: The influence of structure
formation on the cosmic expansion. \emph{Phys. Rev. D} {\bf 80},
083525 (2009) [arXiv:0907.3377 [astro-ph.CO]].

\bibitem{ehlersbuchert:weyl}
J. Ehlers and T. Buchert: On the Newtonian limit of the Weyl tensor.
\emph{Gen. Rel. Grav.} \textbf{41}, 2153 (2009) [arXiv:0907.2645
[gr-qc]].

\bibitem{WS}
A. Wiegand and D. J. Schwarz: Inhomogeneity-induced variance of
cosmological parameters. \emph{Astron. Astrophys.} {\bf 538}, A147
(2012) [arXiv:1109.4142 [astro-ph.CO]].

\bibitem{Stewart}
J. M. Stewart and M. Walker: Perturbations of spacetimes in general
relativity. \emph{Proc. Roy. Soc. Lond. A} {\bf 341}, 49 (1974).


\bibitem{ellis:focus}
G. F. R. Ellis: Inhomogeneity effects in cosmology. \emph{Class.
Quant. Grav.} \textbf{28}, 164001 (2011) [arXiv:1103.2335
[astro-ph.CO]].

\bibitem{clarkson:review}
C. Clarkson, G. F. R. Ellis, J. Larena, and O. Umeh: Does the growth
of structure affect our dynamical models of the Universe? The
averaging, backreaction, and fitting problems in cosmology.
\emph{Rep. Prog. Phys.} \textbf{74}, 112901 (2011) [arXiv:1109.2314
[astro-ph.CO]].

\bibitem{kolb:focus}
E. W. Kolb: Backreaction of inhomogeneities can mimic dark energy.
\emph{Class. Quant. Grav.} \textbf{28}, 164009 (2011).

\bibitem{rasanen:focus}
S. R\"as\"anen: Backreaction: directions of progress. \emph{Class.
Quant. Grav.} \textbf{28}, 164008 (2011) [arXiv:1102.0408
[astro-ph.CO]].

\bibitem{wiltshire:focus}
D. L. Wiltshire: What is dust? Physical foundations of the averaging
problem in cosmology. \emph{Class. Quant. Grav.} \textbf{28}, 164006
(2011) [arXiv:1106.1693 [gr-qc]].

\bibitem{buchert:focus}
T. Buchert: Toward physical cosmology: focus on inhomogeneous
geometry and its non-perturbative effects. \emph{Class. Quant.
Grav.} \textbf{28}, 164007 (2011) [arXiv:1103.2016 [gr-qc]].

\bibitem{roy:perturbations}
X. Roy and T. Buchert: Relativistic cosmological perturbation scheme
on a general background: scalar perturbations for irrotational dust.
\emph{Class. Quant. Grav.} \textbf{29}, 115004 (2012)
[arXiv:1202.5766 [gr-qc]].

\end{thebibliography}
\end{document}